\documentclass[conference]{IEEEtran}
\IEEEoverridecommandlockouts
\usepackage{cite}
\usepackage{amsmath,amssymb,amsfonts}
\usepackage{algorithmic}
\usepackage{graphicx}
\usepackage{textcomp}
\usepackage{xcolor}
\usepackage{comment}
\usepackage{multirow}
\usepackage{array}
\usepackage{listings}
\def\IEEEtitletopspace{20pt}
\newcolumntype{P}[1]{>{\centering\arraybackslash}p{#1}}
\newcolumntype{M}[1]{>{\centering\arraybackslash}m{#1}}
\def\BibTeX{{\rm B\kern-.05em{\sc i\kern-.025em b}\kern-.08em
    T\kern-.1667em\lower.7ex\hbox{E}\kern-.125emX}}
\usepackage{hyperref}
\usepackage{tikz}
\usetikzlibrary{shapes}
\usepackage{xcolor}
\newlength\MAX  

\usepackage{subcaption}
\usepackage{graphicx}
\usepackage{tabularx}
\usepackage{listings}
\usepackage{amsmath}  % Put this in your preamble
\usepackage{array}
\usepackage{titlesec}

\titlespacing{\section}{0pt}{1ex plus 0.5ex minus .2ex}{0.5ex}
\titlespacing{\subsection}{0pt}{0.8ex plus 0.3ex minus .1ex}{0.3ex}

\begin{document}
 \makeatletter
\newcommand{\linebreakand}{%
\end{@IEEEauthorhalign}
\hfill\mbox{}\par
\mbox{}\hfill\begin{@IEEEauthorhalign}
}
\makeatother
\title{How Well Do LLMs Predict Prerequisite Skills? Zero-Shot Comparison to Expert-Defined Concepts \\
}
\author{
	\IEEEauthorblockN{Ngoc Luyen Le\IEEEauthorrefmark{1}\IEEEauthorrefmark{2}\\\textit{ngoc-luyen.le@hds.utc.fr}}
	\and
	\IEEEauthorblockN{Marie-Hélène Abel\IEEEauthorrefmark{1}\\\textit{marie-helene.abel@hds.utc.fr}}
	
	\linebreakand
	\IEEEauthorblockA{
		%\hspace{-1.4cm}
		\IEEEauthorrefmark{1}Université de Technologie de Compiègne, CNRS, Heudiasyc (Heuristics and Diagnosis of Complex Systems),\\ CS 60319 - 60203 Compiègne Cedex, France.
		\\
		%\hspace{-6.5cm}
		\IEEEauthorrefmark{2}Gamaizer, 93340 Le Raincy, France.
		\vspace{-0.5cm}
	}
}
\begin{comment}
\author{\IEEEauthorblockN{1\textsuperscript{st} Given Name Surname}
\IEEEauthorblockA{\textit{dept. name of organization (of Aff.)} \\
\textit{name of organization (of Aff.)}\\
City, Country \\
email address or ORCID}
\and
\IEEEauthorblockN{2\textsuperscript{nd} Given Name Surname}
\IEEEauthorblockA{\textit{dept. name of organization (of Aff.)} \\
\textit{name of organization (of Aff.)}\\
City, Country \\
email address or ORCID}
\and
\IEEEauthorblockN{3\textsuperscript{rd} Given Name Surname}
\IEEEauthorblockA{\textit{dept. name of organization (of Aff.)} \\
\textit{name of organization (of Aff.)}\\
City, Country \\
email address or ORCID}
\and
\IEEEauthorblockN{4\textsuperscript{th} Given Name Surname}
\IEEEauthorblockA{\textit{dept. name of organization (of Aff.)} \\
\textit{name of organization (of Aff.)}\\
City, Country \\
email address or ORCID}
\and
\IEEEauthorblockN{5\textsuperscript{th} Given Name Surname}
\IEEEauthorblockA{\textit{dept. name of organization (of Aff.)} \\
\textit{name of organization (of Aff.)}\\
City, Country \\
email address or ORCID}
\and
\IEEEauthorblockN{6\textsuperscript{th} Given Name Surname}
\IEEEauthorblockA{\textit{dept. name of organization (of Aff.)} \\
\textit{name of organization (of Aff.)}\\
City, Country \\
email address or ORCID}
}
\end{comment}
\maketitle
\begin{abstract}
	Prerequisite skills - foundational competencies required before mastering more advanced concepts - are important for supporting effective learning, assessment, and skill-gap analysis. Traditionally curated by domain experts, these relationships are costly to maintain and difficult to scale. This paper investigates whether large language models (LLMs) can predict prerequisite skills in a zero-shot setting, using only natural language descriptions and without task-specific fine-tuning. We introduce \textit{ESCO-PrereqSkill}, a benchmark dataset constructed from the ESCO taxonomy, comprising 3,196 skills and their expert-defined prerequisite links. Using a standardized prompting strategy, we evaluate 13 state-of-the-art LLMs, including GPT-4, Claude 3, Gemini, LLaMA 4, Qwen2, and DeepSeek, across semantic similarity, BERTScore, and inference latency. Our results show that models such as \textit{LLaMA4-Maverick}, \textit{Claude-3-7-Sonnet}, and \textit{Qwen2-72B} generate predictions that closely align with expert ground truth, demonstrating strong semantic reasoning without supervision. These findings highlight the potential of LLMs to support scalable prerequisite skill modeling for applications in personalized learning, intelligent tutoring, and skill-based recommender systems.

\end{abstract}


\section{Introduction}
The acquisition of knowledge and skills is inherently hierarchical, wherein the mastery of complex competencies is typically contingent upon the prior acquisition of more fundamental, prerequisite skills. Identifying such prerequisite relationships - understanding which skills must logically precede others - is essential for the design of effective pedagogical strategies, structured curriculum development, and adaptive learning systems~\cite{zupan1999learning}. Accurate mapping of these dependencies enhances curriculum sequencing, supports personalized learning interventions, and enables targeted remediation of conceptual gaps~\cite{piech2015deep,chen2018prerequisite}.

Traditionally, the discovery and codification of prerequisite relationships have relied heavily on domain experts, who construct hierarchical frameworks such as Bloom’s Taxonomy~\cite{bloom1956taxonomy}, the European Skills, Competences, Qualifications and Occupations (ESCO)~\footnote{\href{https://esco.ec.europa.eu/en}{https://esco.ec.europa.eu/en}} taxonomy. While these expert-driven frameworks offer structured and pedagogically sound representations, their creation is labor-intensive, expensive to scale, and often lacks the adaptability required in fast-evolving domains. Additionally, manually curated knowledge structures may not capture the level of granularity necessary to support highly personalized and dynamic learning pathways.

Recent advancements in artificial intelligence, particularly the emergence of Large Language Models (LLMs) such as GPT-4~\cite{achiam2023gpt}, Claude~\cite{anthropic_claude3_modelcard_2024}, LLama~\cite{touvron2023llama}, or Gemini~\cite{team2023gemini}, offer a promising alternative. These models, trained on massive and heterogeneous corpora, have demonstrated sophisticated abilities in language understanding, semantic reasoning, and generalization across domains~\cite{wei2022emergent}. Crucially, LLMs may implicitly encode structural and hierarchical relationships among concepts, enabling them to perform zero-shot inference of skill prerequisites - that is, identifying prerequisite relationships without any domain-specific training or fine-tuning. This raises a critical question: To what extent can LLMs infer prerequisite skill relationships directly from their internal knowledge representations, and how do their predictions compare to expert-defined standards? 

In this study, titled ``\textit{How Well Do LLMs Predict Prerequisite Skills? Zero-Shot Comparison to Expert-Defined Concepts}'', we present a rigorous empirical investigation into the zero-shot capabilities of LLMs for prerequisite skill prediction. We introduce a novel benchmark dataset, derived from and validated against the ESCO taxonomy, to serve as a ground truth for evaluating model performance. We systematically assess multiple state-of-the-art LLMs, including LLama, GPT-4, Claude, Gemini, or DeepSeek, using diverse evaluation metrics to quantify their accuracy, consistency, and error patterns. 
Our contributions include defining the zero-shot prerequisite prediction problem in education, synthesizing a benchmark from existing ESCO skill relationships, evaluating several leading LLMs, and analyzing their performance trends and typical errors. This work aims to bridge expert-defined educational structures and AI-driven skill inference, supporting the development of more adaptive and scalable learning technologies.

The remainder of this article is structured as follows: Section \ref{sec_relatedworks} reviews the relevant literature. In Section \ref{sec_methodology}, we describe our methodology, detailing the task formulation and the zero-shot prediction approach for prerequisite skills. Section \ref{sec_experiment_settings} outlines our experimental setup, including the dataset and the LLMs employed. Section \ref{results} presents and discusses the experimental results across different LLMs. Finally, we conclude by examining the implications of our findings and proposing directions for perspectives.

\section{Related work}\label{sec_relatedworks} 
Identifying and modeling prerequisite skill relationships is a long-standing area of research in educational technology and the learning sciences. This field has evolved from expert-crafted frameworks and taxonomies to data-driven computational approaches, and more recently, to methods that leverage the reasoning capabilities of LLMs. This section synthesizes prior work to contextualize our investigation into zero-shot prerequisite skill prediction using LLMs.

Early approaches to modeling educational structure relied on expert-defined frameworks. Bloom’s Taxonomy~\cite{bloom1956taxonomy} and the ACM Computing Curricula~\cite{joint2013computer} provided hierarchical guidelines for learning outcomes and curriculum design. More recent efforts introduced large-scale taxonomies like the ESCO, which links multilingual skills to occupations across Europe. While these structured resources are foundational for semantic consistency and pedagogical soundness, they are difficult to scale and adapt to rapidly changing educational needs~\cite{nan2023personal}.
To address the limitations of expert-driven frameworks, researchers have developed data-driven methods to infer prerequisite relations from educational content. PREREQ~\cite{roy2019inferring} uses topic modeling and a Siamese network to learn dependencies from course descriptions. Pan et al.~\cite{pan2017prerequisite} combine semantic embeddings and course structure to model relations in MOOCs. GKROM~\cite{zhang2025learning} jointly models multiple relation types in a knowledge graph and achieves strong performance through global optimization. These approaches, while effective, often rely on curated datasets or domain-specific features.

The emergence of LLMs such as BERT~\cite{devlin2019bert}, GPT~\cite{achiam2023gpt}, and their more recent successors (as shown in Figure~\ref{fig_01}) has laid the groundwork for generative AI. These models exhibit strong capabilities in semantic reasoning, contextual understanding, and generalization across diverse tasks~\cite{wei2022emergent}. In the educational domain, LLMs are increasingly utilized for intelligent tutoring, automated assessment, feedback generation, and personalized learning recommendations~\cite{di2023retrieval, moore2023empowering}.
Recent research suggests that LLMs can perform structured knowledge extraction - identifying entities and inferring relationships -  even in zero-shot and few-shot scenarios~\cite{yao2025exploring, li2024zero}, often rivaling or surpassing traditional NLP pipelines. However, the specific task of inferring prerequisite relationships among fine-grained educational concepts, particularly in zero-shot settings without domain-specific fine-tuning, remains underexplored.

\begin{figure}[htbp]
	\vspace{-0.2cm}
	\centerline{\includegraphics[width=0.99\linewidth]{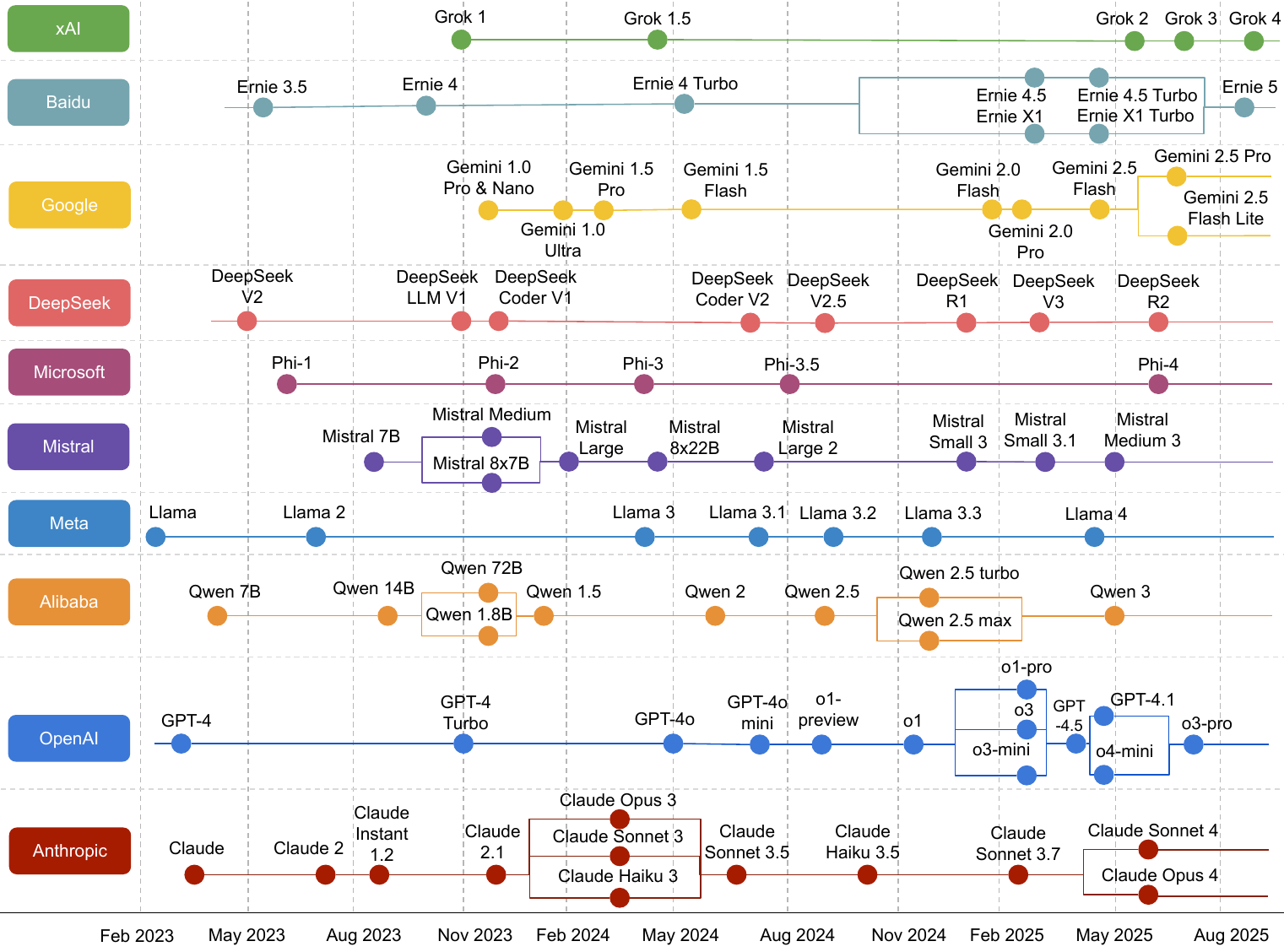}}
	\caption{Timeline of major LLM releases (2023 - Early 2025), showing the rapid evolution of LLMs.}
	\label{fig_01}
	\vspace{-0.5cm}
\end{figure}

Zero-shot prediction - the ability to generalize to unseen tasks without task-specific supervision - has emerged as a key capability of LLMs, enabling them to perform complex reasoning and language understanding tasks. Recent research demonstrates that LLMs can be effectively applied to structured prediction problems such as skill extraction~\cite{herandi2024skill}, recommendation~\cite{liang2024taxonomy}, and natural language understanding~\cite{wang2023zero}. For example, UniMC~\cite{wang2023zero} shows that reformulating tasks into a unified multiple-choice format enhances zero-shot generalization, while TAXREC~\cite{liang2024taxonomy} highlights the benefits of incorporating structured taxonomies into prompts for recommendation. Despite these advances, existing work has primarily focused on general-purpose or high-level tasks. The ability of LLMs to infer prerequisite relationships among fine-grained educational concepts - particularly in a zero-shot setting and relative to expert-defined standards - remains largely unexplored. This study aims to fill that gap by systematically evaluating whether general-purpose LLMs can support prerequisite skill inference without additional task-specific training.
%Zero-shot learning—the ability to generalize to unseen tasks without task-specific supervision—is a core strength of large language models (LLMs). Recent work shows that LLMs can handle structured prediction tasks like skill extraction~\cite{herandi2024skillllm}, recommendation~\cite{liang2025taxrec}, and natural language understanding~\cite{yang2022unimc} without fine-tuning. UniMC~\cite{yang2022unimc} improves generalization through task reformulation, while TAXREC~\cite{liang2025taxrec} leverages structured taxonomies for zero-shot recommendation. However, applying LLMs to infer prerequisite relationships among fine-grained educational skills—especially in alignment with expert-defined standards like ESCO—remains largely unexplored. This study addresses that gap through a systematic evaluation of LLMs' zero-shot capabilities in prerequisite skill inference.

In response to the gaps identified in prior research, this study systematically evaluates the zero-shot capabilities of state-of-the-art LLMs for the task of prerequisite skill prediction. Unlike traditional expert-driven approaches or prior computational methods that rely on domain-specific data and extensive fine-tuning, our approach leverages the general semantic reasoning capabilities of LLMs in a fully automated and scalable way. Central to this effort is a newly synthesized benchmark based on the ESCO taxonomy, providing a structured, expert-validated standard for evaluation. By comparing multiple models using both lexical and semantic metrics, we assess their effectiveness in capturing pedagogical dependencies. %These findings contribute to the broader literature on AI in education and offer actionable insights for integrating LLMs into curriculum design, personalized learning pathways, and intelligent tutoring systems.
The following section presents our methodology, detailing the task formulation, prompting strategy, and evaluation framework used to assess LLM performance on zero-shot prerequisite skill prediction.

\section{Methodology}\label{sec_methodology}
In this section, we introduce the methodology for evaluating how effectively LLMs can predict prerequisite skills without task-specific training. We describe the task formulation and the process for zero-shot skill prediction.

\subsection{Task Formulation}
The task we address involves identifying prerequisite skills for a given target skill or concept using LLMs in a zero-shot setting. A prerequisite skill is defined as a foundational concept or competency that must be acquired prior to effectively learning the target skill. The challenge lies in determining whether LLMs, trained on general, unstructured web-scale corpora, can infer these prerequisite relationships without any fine-tuning or exposure to structured curricula. This task simulates human-like reasoning about learning sequences based solely on natural language understanding.

Formally, let \( T \) denote a target skill, and let \( P = \{p_1, p_2, \dots, p_n\} \) be the set of prerequisite skills as defined by domain experts. The LLM is prompted to generate a predicted set \( \hat{P} = \{\hat{p}_1, \hat{p}_2, \dots, \hat{p}_m\} \) that approximates \( P \). The goal is to ensure that \( \hat{P} \approx P \) both lexically and semantically.

As an example, consider the target skill \( T \)= \text{``$Machine$ $Learning$''} . Experts might define the prerequisite set as \( P  \)= \{\text{``$Linear$ $Algebra$''}, \text{``$Probability$ $Theory$''}, \text{``$Calculus$''}, \text{``$Programming$ $Basics$''}\}. When prompted, the LLM might produce \( \hat{P} \) = \{\text{``$Linear$ $Algebra$''}, \text{``$Probability$ $and$ $Statistics$''}, \text{``$Coding$ $Fundamentals$''}, \text{``$Data$ $Structures$''}\} . Here, ``$Linear$ $Algebra$'' is an exact match, while ``$Probability$ $and$ $Statistics$'' and ``$Coding$ $Fundamentals$'' are semantically related to the ground truth. ``$Data$ $Structures$'' may be relevant but is not part of the expert-defined set.

This problem has practical importance in a range of educational applications. In intelligent tutoring systems, identifying gaps in foundational knowledge allows for adaptive content delivery~\cite{piech2015deep}. In personalized learning platforms, understanding prerequisite relationships supports learner-specific recommendations~\cite{tapalova2022artificial}. Finally, in workforce training scenarios, this task helps define upskilling paths and competencies for job-specific roles~\cite{de2018human}.

Overall, this formulation enables a fair evaluation of LLMs' reasoning abilities in educational contexts without relying on domain-specific training, making it a valuable benchmark for both research and real-world deployment. Building on this foundation, we now describe our approach for zero-shot prerequisite skill prediction, where LLMs are used to infer foundational concepts.

\subsection  {Zero-Shot Prerequisite Skill Prediction}
Zero-shot prediction refers to the ability of a model to generalize to new tasks or domains that were not explicitly seen during training~\cite{xian2018zero}. In the context of prerequisite skill inference, zero-shot prediction leverages LLMs to generate prerequisite concepts for a given target skill without any task-specific fine-tuning. This setting is crucial for evaluating the inherent knowledge representation and reasoning capabilities of general-purpose language models.

\begin{figure}[htbp]
	\vspace{-0.2cm}
	\centerline{\includegraphics[width=0.7\linewidth]{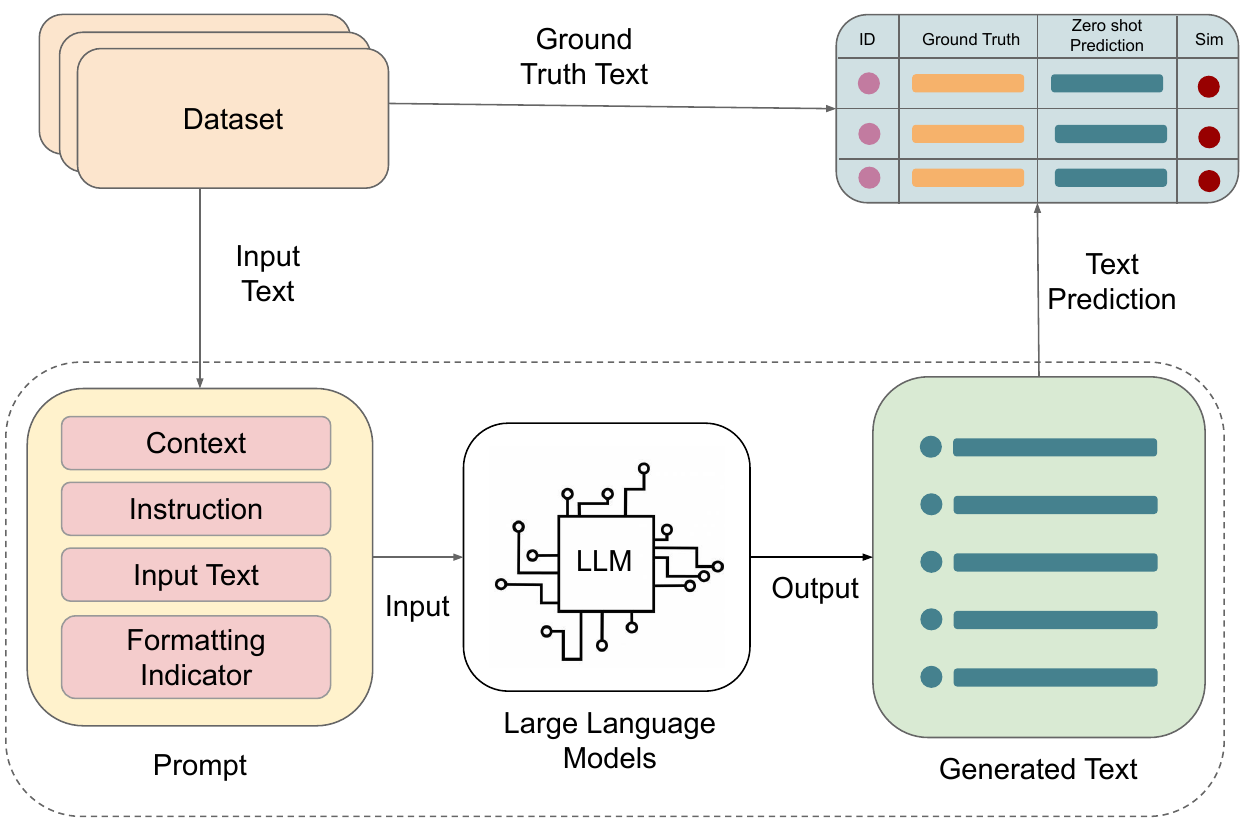}}
	
	\caption{Zero-shot prerequisite skill prediction workflow.}
	\label{fig_02}
	\vspace{-0.7cm}
	
\end{figure}

Modern LLMs are typically pretrained on diverse web-scale corpora and further refined through reinforcement learning with human feedback\cite{ouyang2022training}, making them capable of performing various forms of reasoning\cite{zhang2024chain}. These training methods enable LLMs to exhibit impressive generalization capabilities in both few-shot and zero-shot settings across a wide range of domains\cite{wei2022emergent}. The core premise of zero-shot prerequisite skill prediction is that, through exposure to extensive textual data - including instructional materials, technical documentation, and educational discourse - LLMs may implicitly learn hierarchical and pedagogical relationships among concepts. When prompted with the name and natural language description of a skill, the model is expected to infer and generate a list of foundational concepts that a learner would typically need to understand before acquiring the target skill.

%Our zero-shot prerequisite skill prediction process assesses whether LLMs can infer foundational concepts for a given target skill without any task-specific fine-tuning. As illustrated in Figure~\ref{fig_02}, the workflow begins by selecting a target skill - such as \textit{Machine Learning} - and retrieving its textual description from structured sources like ESCO. This information is embedded into a standardized prompt consisting of four parts: \textit{context}, which introduces the task domain (e.g., skill acquisition); \textit{instruction}, which specifies what the model should do (e.g., list prerequisites); \textit{input text}, containing the skill name and its definition; and a \textit{formatting indicator}, which enforces a clean list-style output. For instance, the input might read: ``\textit{List the essential prerequisite concepts or foundational knowledge areas needed to begin learning the skill: `Machine Learning'. Skill Description: `The ability to design algorithms that learn patterns from data and make predictions'. Respond only with a comma-separated list}''. The prompt is submitted under zero-shot conditions, relying solely on the model’s pretrained knowledge. Predictions are evaluated against expert-defined standards using exact match and semantic similarity to assess the model’s ability to infer pedagogical dependencies.

Our zero-shot prerequisite skill prediction process assesses whether LLMs can infer foundational concepts for a given target skill without any task-specific fine-tuning. As illustrated by the dotted box in Figure~\ref{fig_02}, this part of the workflow begins with selecting a target skill - such as `\textit{Machine Learning}' - and retrieving its textual definition from structured sources like ESCO. This information is embedded into a standardized prompt composed of four components: \textit{context} (introducing the task domain), \textit{instruction} (requesting prerequisite prediction), \textit{input text} (containing the skill name and description), and a \textit{formatting indicator} (specifying the required output format). For example, a prompt might read: ``\textit{List the essential prerequisite concepts or foundational knowledge areas needed to begin learning the skill: `Machine Learning'. Skill Description: `The ability to design algorithms that learn patterns from data and make predictions'. Respond only with a comma-separated list.}'' The prompt is then submitted to the LLM under zero-shot conditions, meaning it relies entirely on its pretrained knowledge. The model's generated list of prerequisite skills is subsequently compared to the expert-defined reference set stored in the benchmark table, which is constructed from the ESCO framework. The evaluation step - detailed in the next section - uses both exact string match and semantic similarity metrics to assess how well the model recovers pedagogical dependencies.

%Our zero-shot prerequisite skill prediction process assesses whether LLMs can infer foundational concepts for a given target skill without any task-specific fine-tuning. As illustrated in Figure~\ref{fig_02}, the process begins with constructing a natural language input that includes the target skill name and its description, typically sourced from structured frameworks. This input is embedded into a standardized prompt and submitted to the LLM in a zero-shot setting. The model then generates a list of predicted prerequisite skills based entirely on its pretrained knowledge. These predictions are evaluated against expert-verified gold standards using both exact match and semantic similarity metrics. Finally, we analyze the performance of multiple LLMs to determine their ability to recover pedagogical dependencies and reason about educational hierarchies using solely natural language.

Ultimately, this zero-shot formulation allows us to test the extent to which general-purpose LLMs exhibit curriculum-like reasoning, making it a strong proxy for evaluating their potential in applications such as personalized education, skill-based recommender systems, and adaptive learning environments. In the following section, we outline the experimental settings used to systematically assess this capability.

\section{Experiment Settings}\label{sec_experiment_settings}
We describe  the experimental setup, covering the dataset, models evaluated, prompt design, and the metrics used for performance assessment.

\subsection{Benchmark Dataset: ESCO-PrereqSkill} 
To ground our experiments in a real-world, expert-validated context, we curated a benchmark dataset named \textit{ESCO-PrereqSkill}, derived from the ESCO - a multilingual, expert-maintained taxonomy that links occupational roles to required skills, including explicitly defined prerequisite relationships.

\begin{table}[ht]
	\centering
	\caption{Example from the ESCO-PrereqSkill Dataset.}\label{table_01}
	\begin{tabular}{|m{1.5cm}|m{6.5cm}|}
		\hline
		\textbf{Skill Name} & Supervise correctional procedures \\\hline
		\textbf{Skill ID} & \texttt{00064735-8fad-454b-90c7-ed858cc993f2} \\\hline
		\textbf{Description} & Supervise the operations of a correctional facility or other correctional procedures, ensuring that they are compliant with legal regulations, and ensure that the staff complies with regulations, and aim to improve the facility's efficiency and safety. \\\hline
		\textbf{Ground Truth Prerequisites} & Correctional procedures,\newline Legal regulations in corrections,\newline Staff management,\newline Safety protocols.\\\hline
	\end{tabular}
	\vspace{-0.2cm}
\end{table}

We focused on the `skills/competences' pillar of the ESCO taxonomy. Target skills were selected based on three criteria:  (i) the availability of a clear and descriptive skill name; (ii) the presence of a textual description providing sufficient semantic context; and (iii) the existence of linked prerequisite skills, identified through relational properties such as the inverse of \texttt{isEssentialSkillFor}. These relationships were used to extract ground truth prerequisite lists, which serve as the reference outputs for model evaluation.

Each instance in the benchmark includes the skill ID, name, description, and the corresponding list of expert-defined prerequisite skills, as illustrated in Table~\ref{table_01}. The final dataset contains 3,196 skills, each with one or more prerequisite skills, and is publicly available\footnote{\href{https://github.com/lengocluyen/ESCO-PrereqSkill}{https://github.com/lengocluyen/ESCO-PrereqSkill}} as a test set for fair zero-shot evaluation.

\subsection{LLMs Selection}

We selected a diverse set of 13 LLMs spanning various architectural families, model sizes, and providers. This selection includes both open and commercial models, enabling a robust comparison of zero-shot prerequisite skill prediction across a wide spectrum of capabilities—from frontier proprietary systems to efficient publicly accessible variants. The 13 models evaluated in this study are categorized according to their respective organizations as follows:

\textit{OpenAI}: \textit{gpt-4.5-preview} (February 2025) and \textit{gpt-4o} (May 2024) offer enhanced reasoning and faster, multimodal inference. \textit{o1-mini}, released in early 2025, is a compact variant optimized for low-latency applications.

\textit{Anthropic}: \textit{claude-3.5-haiku} (June 2024), \textit{claude-3.7-sonnet} (Feb 2025), and \textit{claude-3-opus} (Mar 2024) are Claude 3 models optimized for safety and alignment. Haiku prioritizes speed, Sonnet balances performance, and Opus offers the strongest reasoning.

\textit{Google}: \textit{gemini-2.0-flash} (February 2025) and \textit{gemini-2.5-pro-preview} (March 2025) are part of the Gemini series, using Mixture-of-Experts architectures to support efficient multimodal reasoning with sparse activation and high throughput.

\textit{Meta}: \textit{llama4-maverick} and \textit{llama4-scout} (both released April 2025) are LLaMA 4 variants designed for open deployment. Maverick targets high performance on reasoning benchmarks, Scout prioritizes low latency and smaller scale.

\textit{DeepSeek AI}: \textit{deepseek-v3} (December 2024) and \textit{deepseek-r1} (March 2025) are transformer models trained for multilingual and code-heavy tasks, with efficient long-context support and strong zero-shot generalization.

\textit{Alibaba}: \textit{Qwen2-72B} (June 2024) is a large transformer model with 128K context and instruction tuning for reasoning, summarization, and dialogue.

All models were accessed via official APIs or inference endpoints. No additional tuning or adaptation was applied, preserving the integrity of the zero-shot experimental setting.

\subsection{Zero-Shot Generation Protocol}

To ensure consistency and comparability, we employed a unified zero-shot generation protocol across all models. Each LLM was queried using a standardized prompt that framed the task as prerequisite skill identification, grounded in the target skill’s name and description.

\begin{figure}[htbp]
	\vspace{-0.2cm}
	\centerline{\includegraphics[width=0.7\linewidth]{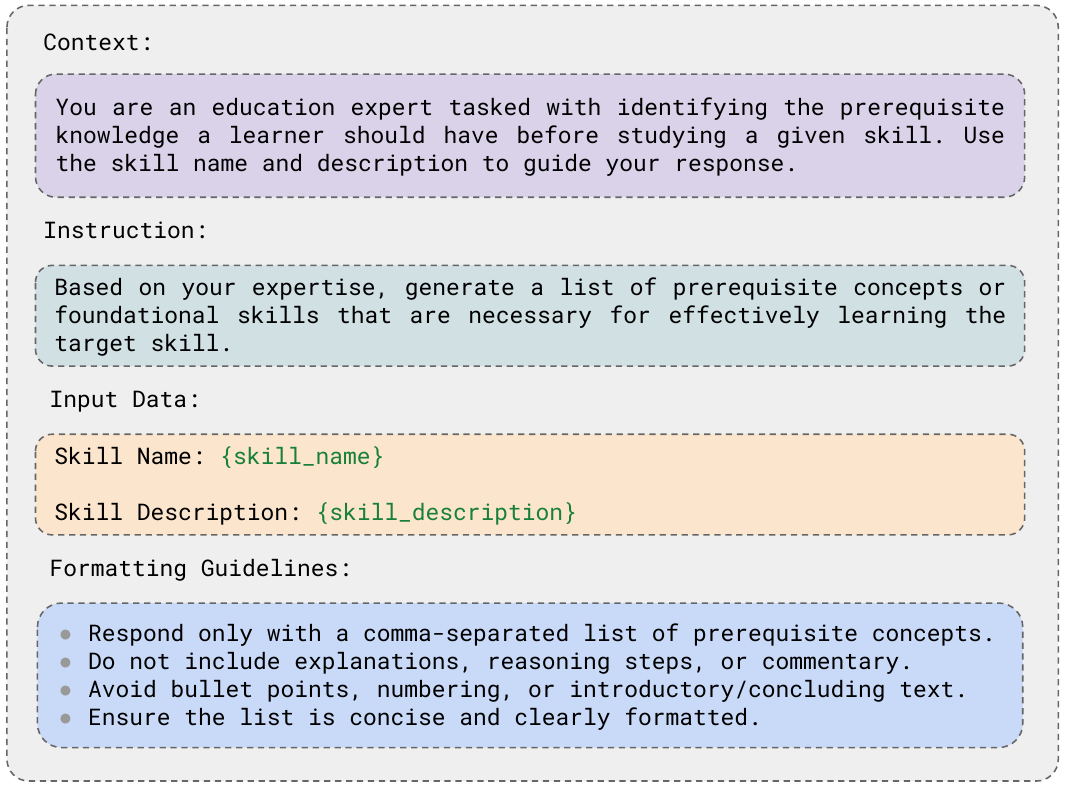}}
	
	\caption{Zero-shot prompt template.}
	\label{fig_03}
	\vspace{-0.3cm}
	
\end{figure}

The prompt was carefully designed to elicit concise, structured, and comparable outputs across models, as shown in Figure~\ref{fig_03}. Each model receives a standardized instruction to act as an education expert and return a comma-separated list of prerequisite concepts without explanations or additional formatting. To prepare the outputs for evaluation, responses were parsed into list structures and normalized through lowercasing, whitespace trimming, and punctuation removal. These preprocessing steps minimize surface-level formatting differences and enhance the robustness of both lexical and semantic comparisons.

\subsection{Evaluation Metrics}

The open-ended and generative nature of the task necessitates semantic evaluation metrics, as exact string matching is insufficient for capturing conceptual alignment. We employed two complementary evaluation strategies:

\begin{figure*}[th]
	\vspace{-0.2cm}
	\centerline{\includegraphics[width=0.850\linewidth]{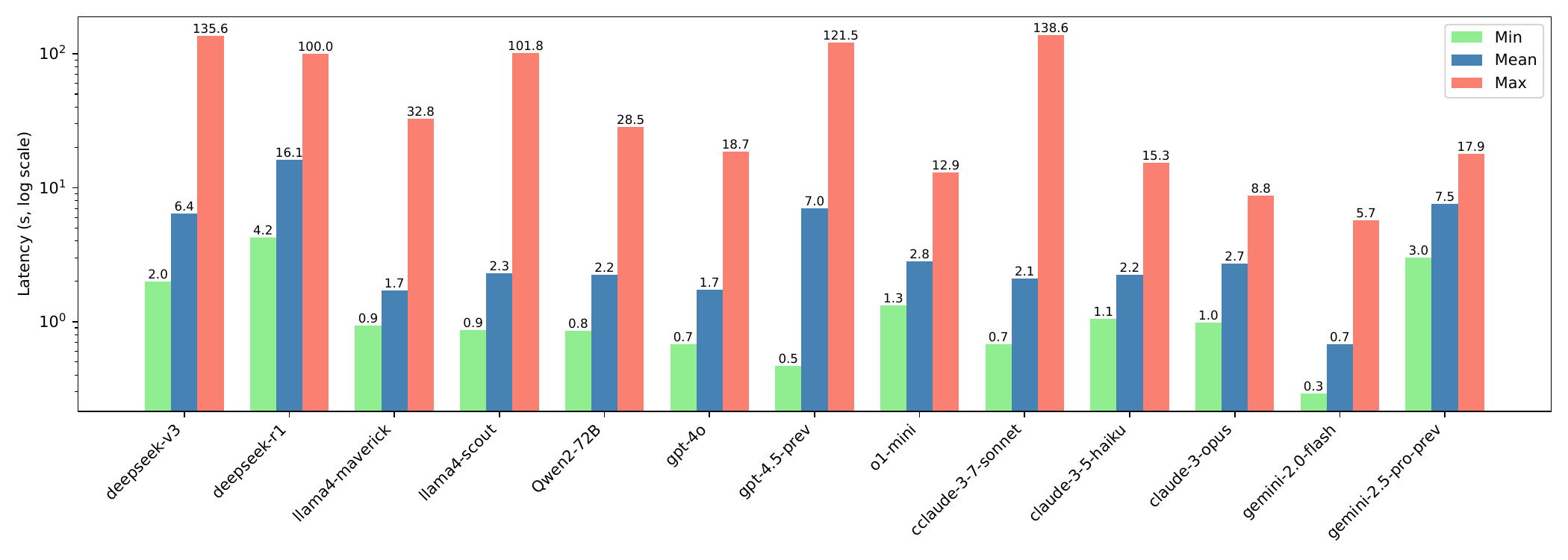}}
	
	\vspace{-0.4cm}
	\caption{Comparison of minimum, mean, and maximum inference latency (in seconds, log scale) for different LLMs.}
	
	\label{fig_04}
	
	\vspace{-0.4cm}
\end{figure*}

\textit{Semantic Similarity:}  
This metric evaluates the overall semantic alignment between the predicted set \( \hat{P} \) and the expert-defined set \( P \). Each set is concatenated into a single string and embedded using Sentence-BERT~\cite{reimers-2019-sentence-bert}. The similarity score is computed as the cosine similarity between the two embeddings: \( \text{Sim}_{\text{sem}}(\hat{P}, P) = \frac{f(\hat{P}) \cdot f(P)}{\|f(\hat{P})\| \|f(P)\|} \), where \( f(\cdot) \) denotes the Sentence-BERT embedding function.

\textit{BERTScore (Precision, Recall, F1):}  
BERTScore~\cite{zhang2019bertscore} provides token-level comparison using contextual embeddings. Recall is computed as \( R_{\text{BERT}} = \frac{1}{|P|} \sum_{p_i \in P} \max_{\hat{p}_j \in \hat{P}} \mathbf{p}_i^\top \hat{\mathbf{p}}_j \), and precision as \( P_{\text{BERT}} = \frac{1}{|\hat{P}|} \sum_{\hat{p}_j \in \hat{P}} \max_{p_i \in P} \mathbf{p}_i^\top \hat{\mathbf{p}}_j \). Each skill in the reference set \( P \) is matched to its most similar predicted counterpart in \( \hat{P} \), and vice versa. The final F1 score is given by \( F1_{\text{BERT}} = \frac{2 P_{\text{BERT}} R_{\text{BERT}}}{P_{\text{BERT}} + R_{\text{BERT}}} \).

All metrics were computed over the full test set and averaged across examples. This dual-metric evaluation captures both specific and general accuracy, enabling fine-grained performance comparisons. The next section presents experimental results, detailing model performance across lexical and semantic dimensions.

\section{Experimental Results}\label{results}

We evaluate 13 leading LLMs on zero-shot prerequisite skill prediction, using expert-curated relationships from the ESCO taxonomy as ground truth. This section presents quantitative results and qualitative insights into model behavior, focusing on semantic alignment and inference efficiency.

\begin{table}[ht]
	\centering
	\caption{Model performance on zero-shot prerequisite skill prediction using semantic similarity and BERTScore metrics.}
	\begin{tabular}{|m{2.4cm}|P{1.3cm}|m{1.0cm}|m{1.0cm}|M{1.0cm}|}
		\hline \multicolumn{1}{|c|}{\textbf{Model}}
		& \centering\textbf{$\text{Sim}_{\text{sem}}$} & \centering\textbf{$P_{\text{BERT}}$} & \centering\textbf{$R_{\text{BERT}}$} & \textbf{$F1_{\text{BERT}}$} \\
		\hline
		claude-3-7-sonnet        & \textbf{0.7140} & 0.7773 & 0.8610 & 0.8167 \\\hline
		claude-3-5-haiku         & 0.6380 & 0.7733 & 0.8536 & 0.8111 \\\hline
		claude-3-opus            & 0.6988 & 0.7831 & 0.8600 & 0.8194 \\\hline
		deepseek-r1              & 0.6844 & 0.7727 & 0.8597 & 0.8136 \\\hline
		deepseek-v3              & 0.7011 & 0.7803 & 0.8613 & 0.8185 \\\hline
		gemini-2.0-flash         & 0.6049 & 0.7959 & 0.8543 & 0.8238 \\\hline
		gemini-2.5-pro-prev   & 0.6185 & 0.7861 & 0.8547 & 0.8187 \\\hline
		gpt-4.5-prev          & 0.6988 & 0.7771 & 0.8594 & 0.8159 \\\hline
		gpt-4o                   & 0.6806 & 0.7905 & 0.8584 & 0.8227 \\\hline
		llama4-maverick          & 0.7108 & \textbf{0.8071} & \textbf{0.8650} & \textbf{0.8347} \\\hline
		llama4-scout             & 0.7117 & 0.7926 & 0.8622 & 0.8256 \\\hline
		o1-mini                  & 0.6745 & 0.7872 & 0.8594 & 0.8214 \\\hline
		Qwen2-72B                & 0.7005 & 0.7934 & 0.8625 & 0.8262 \\
		\hline
	\end{tabular}
	\label{table_02}
	\vspace{-0.3cm}
\end{table}

\subsection{Prediction Performance}

Table~\ref{table_02} reports semantic similarity and \textit{BERTScore} metrics across all models. Several LLMs demonstrate strong alignment with expert-defined prerequisite concepts. \textit{LLaMA4-Maverick} achieved the highest $F1_{\text{BERT}}$ score (0.8347), followed closely by \textit{Qwen2-72B} (0.8262), \textit{LLaMA4-Scout} (0.8256), and \textit{Gemini-2.0-Flash} (0.8238). These results suggest that the top-performing models can recover meaningful prerequisite structures purely from textual prompts, even without domain-specific fine-tuning.

However, discrepancies between BERTScore and semantic similarity reveal differences in model reasoning. For example, \textit{Gemini-2.0-Flash} achieves a high  $F1_{\text{BERT}}$ score but lower semantic similarity (0.6049), suggesting surface-level lexical matching rather than true conceptual understanding. In contrast, \textit{Claude-3-7-Sonnet} and \textit{LLaMA4-Scout} achieve a better balance between lexical overlap and semantic depth, indicating stronger internal representations of prerequisite logic.

\subsection{Inference Efficiency}

To assess practical deployability, we also analyzed latency across models (see Figure~\ref{fig_04}). Some models, such as \textit{Gemini-2.0-Flash}, \textit{GPT-4o}, and \textit{LLaMA4-Maverick}, delivered predictions in under 2 seconds on average, making them viable for interactive educational systems. Others, such as \textit{DeepSeek-R1} (16.09s) and \textit{GPT-4.5-Preview} (7.04s), demonstrated slower response times despite competitive output quality.

Importantly, several top-performing models - \textit{Claude-3-Opus}, \textit{LLaMA4-Maverick}, and \textit{Qwen2-72B} - achieve both high accuracy and low-to-moderate latency, showing that good prerequisite reasoning doesn’t have to slow down inference.

\subsection{Interpreting Model Capabilities}

Taken together, the results provide a positive answer to the guiding question posed in the introduction. Modern LLMs can, to a significant extent, infer prerequisite skill relationships from skill names and descriptions alone. High-performing models exhibit internalized structural knowledge that aligns with expert-defined competence hierarchies. This emergent capability is especially notable given that models operate in a zero-shot setting, with no exposure to labeled training examples or domain-specific instructional frameworks.

While some models rely more heavily on surface-level lexical patterns, others demonstrate deeper reasoning abilities that reflect structured logic commonly found in human-authored competence models. These differences highlight the importance of interpreting model capacities not just by accuracy metrics, but by their ability to generalize pedagogical relationships across domains. Such interpretability is crucial for determining whether LLMs can reliably support downstream applications in skill-gap analysis, personalized learning, and competence-based workforce development.

In summary, our findings support the feasibility of using general-purpose LLMs for skill-based competence management - without requiring task-specific supervision. Future work may explore hybrid approaches that combine structured ontologies with generative reasoning to enhance both the reliability and transparency of model outputs.

\section{Conclusion and Perspective}\label{sec_conclusion}
This study examined whether LLMs can predict prerequisite skills in a zero-shot setting, without fine-tuning or access to structured competence frameworks. Leveraging expert-defined relationships from the ESCO taxonomy, we introduced a new benchmark and evaluated 13 state-of-the-art LLMs across semantic and lexical metrics, as well as inference latency.
Our results demonstrate that several general-purpose LLMs - particularly \textit{LLaMA4-Maverick}, \textit{Claude-3-7-Sonnet}, and \textit{Qwen2-72B} - are capable of producing prerequisite skill predictions that closely align with expert-verified concept structures. These models exhibit strong semantic reasoning and competitive performance, even when operating under strict zero-shot constraints.
Moreover, our latency analysis reveals that many high-performing models are also suitable for real-time educational applications, offering a practical path toward scalable, AI-driven curriculum design and personalized learning tools.
This work demonstrates the potential of LLMs for educational reasoning and lays the groundwork for future research in adaptive learning and hybrid neuro-symbolic approaches. Next steps include exploring few-shot settings with the ESCO-PrereqSkill dataset, fine-tuning for better alignment, developing hybrid models, and evaluating robustness to prompt variation and domain shifts.

\section*{Acknowledgments}
We warmly thank the Ikigai consortium led by the association Games for Citizens, the company Gamaizer, as well as the FORTEIM project (winner of the AMI CMA France 2030 call for projects), for their support and collaboration. Their contributions have provided significant added value to the completion of this research.
\bibliographystyle{ieeetr}
\bibliography{references}

\end{document}